\documentclass[12pt,preprint]{aastex} 
\usepackage{graphicx,emulateapj5,apjfonts} 
\usepackage{onecolfloat} 
\usepackage{epsf} 
 
\shortauthors{Wolf et al.} 
\shorttitle{GEMS: UV luminosity density at $z\sim 0.7$} 
\slugcomment{{\sc ApJ, re-submitted after replying to referee's report: }
  \today } 
 
\begin{document} 
 
\def\head{ 
 
\title{GEMS: which galaxies dominate the $z \sim 0.7$ ultraviolet 
  luminosity density?} 
 
\author{C. Wolf$^1$, E. F. Bell$^2$,  D. H. McIntosh$^6$, H.-W. Rix$^2$, 
	M. Barden$^2$, S. V. W. Beckwith$^{3,4}$, \\ 
	A. Borch$^2$, J. A. R. Caldwell$^3$, B. H\"au\ss ler$^2$, 
	C. Heymans$^2$, K. Jahnke$^5$, S. Jogee$^3$, \\ 
	K. Meisenheimer$^2$, C. Y. Peng$^7$, S. F. S\'anchez$^5$, 
	R. S. Somerville$^3$, L. Wisotzki$^5$} 
\affil{$^1$ Department of Physics, Denys Wilkinson Bldg., 
            University of Oxford, Keble Road, Oxford, OX1 3RH, U.K. \\ 
       $^2$ Max-Planck-Institut f\"ur Astronomie, K\"onigstuhl 17, 
            D-69117 Heidelberg, Germany \\ 
       $^3$ Space Telescope Science Institute, 
            3700 San Martin Drive, Baltimore, MD 21218, U.S.A. \\ 
       $^4$ Johns Hopkins University, Charles and 4th Street, 
            Baltimore, MD 21218, U.S.A. \\ 
       $^5$ Astrophysikalisches Institut Potsdam,  
            An der Sternwarte 16, D-14482 Potsdam, Germany \\ 
       $^6$ Dept. of Astronomy, University of Massachusetts,  
            710 North Pleasant Street, Amherst, MA 01003, U.S.A. \\ 
       $^7$ Steward Observatory, University of Arizona, 
            933 North Cherry Avenue, Tucson, AZ 85721, U.S.A. \\ 
} 
 
\begin{abstract} 
We combine high-resolution images from GEMS with redshifts and spectral  
energy distributions from COMBO-17 to explore the morphological types of  
galaxies that dominate the $z \sim 0.7$ UV luminosity density. We analysed 
rest-frame 280~nm and $V$-band luminosities of 1483 galaxies with $0.65<z<
0.75$, combining these with visual morphological classifications from 
F850LP images (approximately rest-frame $V$-band) taken with HST/ACS on 
the Extended Chandra Deep Field South. We derive UV luminosity functions 
and $j_{280}$ luminosity densities for spheroid-dominated galaxies, 
spiral galaxies, Magellanic irregulars, and clearly-interacting  
galaxies with morphologies suggestive of ongoing  
major mergers. We check the reliability  
of GEMS morphologies against the deeper GOODS images and quantify an  
incompleteness of the GEMS merger identification at the faint end.  
We derive the fractions of the global UV luminosity density $j_{280}$  
originating from the galaxy types, and find that spiral galaxies and  
Magellanic irregulars dominate with about 40\% each. Interacting and merging 
galaxies account for roughly 20\% of $j_{280}$, while the contribution  
of early types is negligible. 
 
These results imply that the strong decline in the UV luminosity density 
of the Universe observed from $z\sim1$ until today is dominated by the 
decreasing UV luminosities of normal spiral galaxies, accompanied by the 
migration of UV-luminous star formation in irregular galaxies to systems 
of progressively lower mass and luminosity. These conclusions 
suggest that major merger-driven star formation cannot 
dominate the declining cosmic star formation rate, unless 
major mergers are {\it both} substantially more obscured than  
intensely star-forming spiral galaxies {\it and} the decline  
in observed cosmic star formation rate is substantially stronger 
than the already precipitous decline in uncorrected UV luminosity density.  
\end{abstract} 
 
\keywords{Surveys  --  Galaxies: evolution  --  Galaxies: general   
   --  Galaxies: luminosity function  --  Galaxies: morphology} 
}%%%end head 
\twocolumn[\head]

\section{Introduction} 
 
\begin{figure*} 
\centering 
\hbox{ 
\includegraphics[clip,width=0.1\hsize]{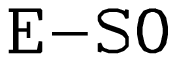} 
\includegraphics[clip,width=0.2\hsize]{f1ab.eps} 
\includegraphics[clip,width=0.2\hsize]{f1ac.eps} 
\includegraphics[clip,width=0.2\hsize]{f1ad.eps} 
\includegraphics[clip,width=0.2\hsize]{f1ae.eps}} 
\hbox{ 
\includegraphics[clip,width=0.1\hsize]{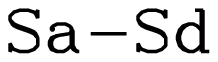} 
\includegraphics[clip,width=0.2\hsize]{f1bb.eps} 
\includegraphics[clip,width=0.2\hsize]{f1bc.eps} 
\includegraphics[clip,width=0.2\hsize]{f1bd.eps} 
\includegraphics[clip,width=0.2\hsize]{f1be.eps}} 
\hbox{ 
\includegraphics[clip,width=0.1\hsize]{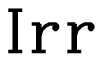} 
\includegraphics[clip,width=0.2\hsize]{f1cb.eps} 
\includegraphics[clip,width=0.2\hsize]{f1cc.eps} 
\includegraphics[clip,width=0.2\hsize]{f1cd.eps} 
\includegraphics[clip,width=0.2\hsize]{f1ce.eps}} 
\hbox{ 
\includegraphics[clip,width=0.1\hsize]{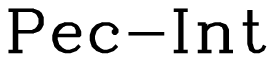} 
\includegraphics[clip,width=0.2\hsize]{f1db.eps} 
\includegraphics[clip,width=0.2\hsize]{f1dc.eps} 
\includegraphics[clip,width=0.2\hsize]{f1dd.eps} 
\includegraphics[clip,width=0.2\hsize]{f1de.eps}} 
\caption{HST/ACS F850LP example images of the four broad type bins 
types; each class has four examples.  
Panels measure $6\arcsec \times 6\arcsec$, 
corresponding to roughly 45\,kpc$ \times $45\,kpc.  
{\it From top row to bottom row:} E--S0 --- Sa--Sd --- Irr --- Pec/Int.} 
\label{examples} 
\end{figure*} 
 
It is well-established that the UV luminosity density integrated over 
the galaxy population drops from $z\ga 1$ to the 
present day \citep{Mad96,Lil96,Cow96,Mad98}, amounting to a factor of  
three to six decline between $z=0.7$ and $z=0$ at $\lambda_{rest} = 280$~nm 
%% \citep{Lil96,Cow96,Wil02,Wolf03} 
(Lilly et al. 1996; Cowie et al. 1996; Wilson et al. 2002; Wolf et al.  
2003)\footnote{In this paper we assume $(\Omega_m,\Omega_\Lambda)= 
(0.3,0.7)$ and $H_0=h \times 100$\,km/sec.}.  
Because UV light is emitted by young, massive stars, this decline is  
conceptually linked to a dramatic decline in the global cosmic star  
formation rate (SFR).  Despite the fact that most UV photons are absorbed by  
dust \cite{WH96}, observations of less dust-sensitive SFR indicators  
support this interpretation (Flores et al. 1999; Haarsma et al. 2000;  
Yan et al. 1999), although they do not reach as far down the luminosity 
function.  The goal of this paper is to explore the physical 
processes driving the evolution of the cosmic SFR by  
investigating the contribution of different galaxy morphological types 
to the cosmic-averaged UV luminosity density at $z \sim 0.7$, 
when the universe was half of its present age. 
 
The physics that drives this decline remains unclear.   
Observations of damped Lyman-$\alpha$ systems and local 
{\sc Hi} surveys indicate that the cosmic density of cold 
neutral gas also declines dramatically from 
$z\sim1$ to the present (e.g. Storrie-Lombardi, McMahon, \& Irwin 
1996). Some of the decline in cold gas is simply 
because a significant fraction of the available baryons become locked 
up in long-lived stars (Pei, Fall, \& Hauser 1999).  The hierarchical 
formation of massive structures at $z\la2$ also leads to a larger 
fraction of gas being in a shocked warm/hot phase, with long cooling 
times \citep[e.g. ][]{CO99}, contributing further 
to the declining cosmic cold neutral gas density.  
If the efficiency of star formation 
is linked to the surface density of gaseous disks, as suggested by 
local observations (Kennicutt 1998), and if gas disks are denser at 
high redshift, this alone may dominate the declining cosmic SFR  
(Somerville, Primack \& Faber 2001).  Yet, galaxy interactions are 
known to trigger efficient bursts of star formation \cite[e.g.][]{SM96};
it is likely that a decline in the merger rate (Le F\`evre et al. 2000; 
Carlberg et al. 2000; Patton et al. 2002; Conselice et al. 2003; Bundy 
et al. 2004; although see Lin et al. 2004 for a different conclusion) 
will also contribute to the declining SFR over this period. 
 
The characteristic rise and decline of the cosmic SFR 
is a generic prediction of cosmological simulations of galaxy formation  
in the Cold Dark Matter framework \citep{bau98,spf,nag2000,sh03,hs03}.  
But the location of the SFR maximum and the rate of the decline 
depend on the input physics, especially the treatment of star  
formation and feedback. Hence, observational tests of star formation 
in different types of galaxies are very interesting for constraining 
these model ingredients. 
 
\begin{figure} 
\centering 
\includegraphics[clip,angle=270,width=0.7\hsize]{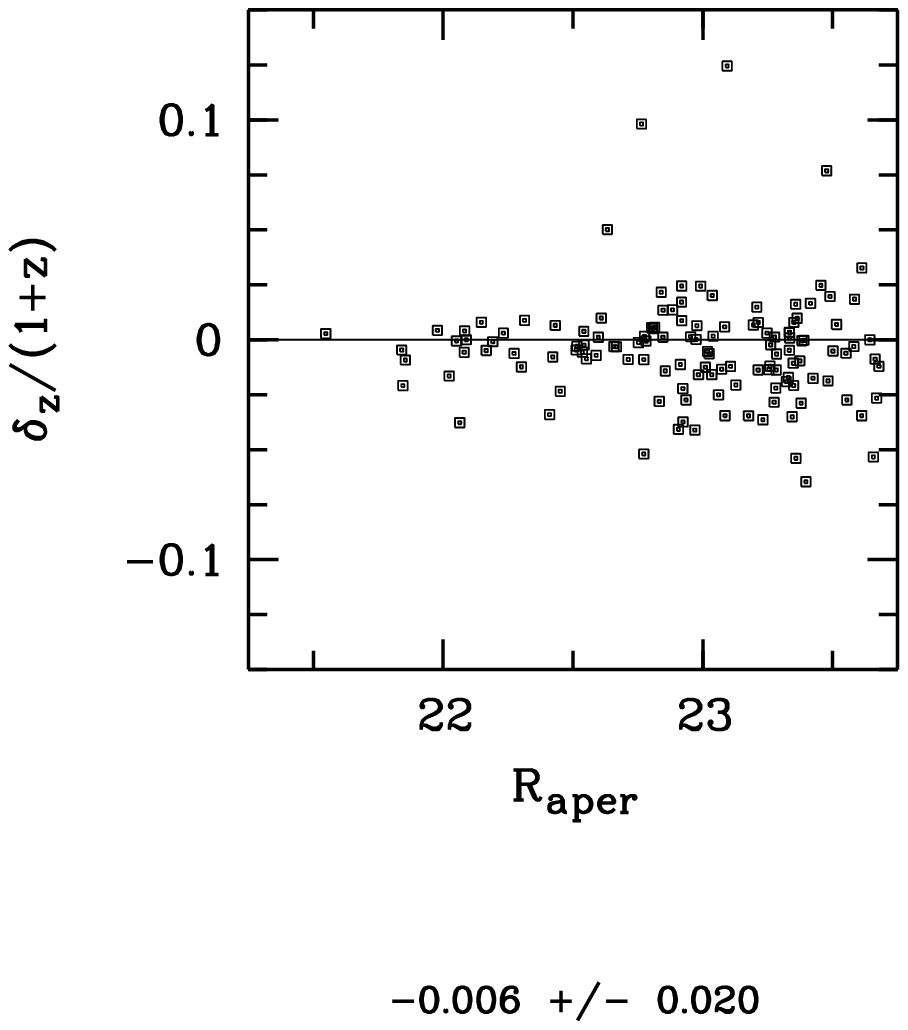} 
\caption{Redshift errors: The VVDS survey provides redshifts with $>95$\%
reliability for 152 of the 1483 galaxies in our sample. Three outliers can
be seen with redshift errors on the order of $\delta_z\sim 0.1$. The bulk
of the sample (98\%) have a mean offset from the spectroscopic redshift of
$\langle \delta_z/(1+z) \rangle = -0.006$ and an RMS scatter of 0.020 in 
terms of $\delta_z/(1+z)$.}
\label{zz} 
\end{figure}

\begin{figure} 
\centering 
\includegraphics[clip,angle=0,width=0.7\hsize]{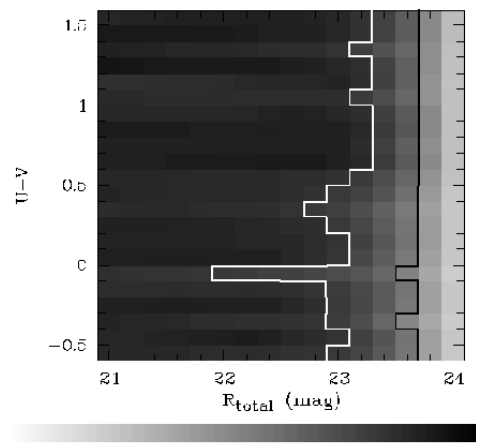} 
\caption{Sample completeness: Monte-Carlo simulations of the COMBO-17 
survey with classification and redshift estimation yield this map of
completeness for galaxies at $0.65<z<0.75$. Dark grey denotes high 
completeness, light grey resembles low completeness. Lines are drawn
at 50\% completeness (black) and 90\% (white). At $R<23$ the sample is
more than 90\% complete, at $R=23.7$ still $\sim 50$\% of the galaxies
are included. At $R>24$ no galaxy is included, because of the $R_{\rm 
aper}<24$ selection cut.}
\label{cmap} 
\end{figure} 
 
In this paper, we present progress towards pinning down the underlying  
physics responsible for the decline in the cosmic SFR, through 
an investigation of the contribution of galaxies with 
different morphological types to the UV luminosity density 
at $z\sim 0.7$. For this investigation we combine high 
spatial resolution imaging from the Advanced Camera for Surveys (ACS) 
on the Hubble Space Telescope --- taken from the GEMS  
\cite[Galaxy Evolution from Morphology and SEDs; ][]{Rix04} 
survey --- with accurate photometric 
redshifts and rest-frame UV/optical luminosities and colors from 
COMBO-17 \citep[Classifying Objects by Medium-Band Observations in 17 
filters; ][]{Wolf01,Wolf04}.  In particular, we will assess the UV  
luminosity contribution of manifestly interacting or morphologically  
peculiar galaxies at $z\sim 0.7$ compared to, e.g., luminous spirals.  
We restrict our attention to a redshift slice at $0.65<z<0.75$, where  
the ACS F850LP band corresponds roughly to the rest-frame $V$-band, 
in order to bypass uncertainties due to morphological $K$-corrections 
and to minimize any redshift-dependent selection effects, e.g. due to 
cosmological surface brightness dimming. Our sample contains almost 
1500 galaxies with COMBO-17 redshifts and visually-classified 
morphological types. A more detailed study over the full redshift 
range probed by the GEMS sample will be presented in future work. 
 
The rest-frame UV luminosity is an important tracer for star formation  
all the way to $z\ga 6$ (e.g. Sullivan et al. 2000; Giavalisco et  
al. 2002; Bouwens et al. 2003; Bunker et al. 2004). Of course, a fuller  
understanding of how the SFR density evolves will have to incorporate  
complementary constraints on the bolometric luminosity of the youngest 
stellar population, such as H$\alpha$, IR, or radio luminosities.  
In order to directly constrain the amount of dust-obscured star 
formation in our sample, a first analysis of Spitzer 24$\micron$ data 
taken by the MIPS instrument team (Rieke et al. 2004) on the Chandra 
Deep Field South is presented in a companion paper (Bell et al. 2005). 
However, the advantage of the UV analysis presented here is that  
observations of rest-frame UV probe much further down the luminosity 
function than all other SFR indicators at $z \ga 0.5$, meaning that 
the total UV luminosity density is relatively reliably constrained.  
Finally, this study will reveal which galaxies are responsible for the  
changing UV ionising background at $z\la 1$, because galaxies with  
large contributions to the 280~nm luminosity density are likely to make 
correspondingly large contributions to the cosmic luminosity density in 
the ionizing background at $\lambda<91$~nm. 
 
The structure of this paper is as follows. In \S~2 we describe the 
data sets and in \S~3 our morphological 
classification. In \S~4 we discuss the distribution of UV/optical 
galaxy colors, present UV luminosity functions by morphological type,  
and assess the contribution of different types to the rest-frame UV  
luminosity density. In \S~5 we discuss our results in the light of  
previous results and theoretical predictions, summarizing in \S~6.

\begin{figure} 
\centering 
\includegraphics[clip,angle=270,width=0.7\hsize]{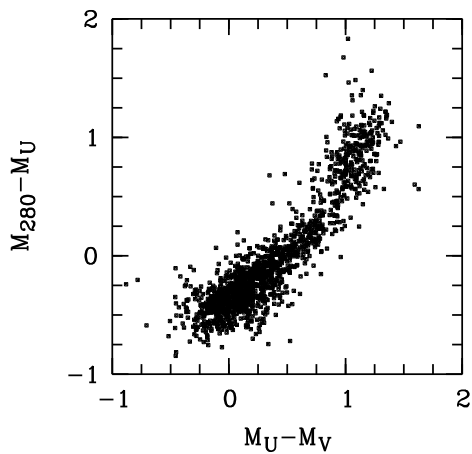} 
\caption{Rest-frame colors: The sample galaxies form almost a 
one-parameter family in near-UV-optical rest-frame colors. The 
bimodality can be seen as well.}
\label{colcol} 
\end{figure}

\section{Data} 
 
The COMBO-17 survey has measured accurate photometric redshifts for 
more than 10,000 galaxies in the Extended Chandra Deep Field South 
\citep[E-CDFS;][]{Wolf04}. The redshifts are estimated using data
from 5 broad and 12 medium bands. We use seeing-adaptive aperture 
photometry measuring the same physical fraction of a galaxy in each
passband and concentrate the aperture on the brighter central parts.
This way we ensure a maximum signal-to-noise ratio for recovering
the SED and redshift of any object (see Wolf et al. 2004 for all 
details). Detailed comparisons with spectroscopic redshifts and 
simulations have suggested the mean error of COMBO-17 redshifts to be 
$\sigma_z/(1+z) \sim 0.02$ for galaxies with $R<23$ and $z<1.2$. COMBO-17
has also identified type-1 AGN when the contribution of AGN light to the 
SED was sufficient to leave recognizable traces in the 17-band data set.

The GEMS survey has covered almost the whole COMBO-17 area on the E-CDFS 
with HST/ACS observations on an area of $\sim 800\sq\arcmin$   % AREA 796.4  
\citep[Caldwell et al. in preparation,][]{Rix04}. Altogether, 77 HST/ACS 
pointings were observed with an exposure time of one orbit each in the  
F606W and F850LP passbands. The central 1/5th of the area is incorporated  
from the first epoch of GOODS imaging (Giavalisco et al. 2004)\footnote{
Later, we compare morphological classifications from first-epoch GOODS 
imaging with the full-depth GOODS imaging to explore how the results from 
GEMS may be systematically biased by the shallower imaging data.}. For  
galaxy classification, the images are approximately surface brightness  
limited, at a limiting average surface brightness within the half-light  
radius of $z_{AB}\sim 24$~mag/$\sq\arcsec$ in the F850LP passband.

For this paper, we have selected all galaxies from the overlapping area 
of COMBO-17 and GEMS, which have no measurable type-1 AGN contribution,
photometric redshifts of $0.65<z<0.75$, and aperture magnitudes of $R
_{aper}<24$. At $R_{aper}<24$ our photometric redshifts are well-behaved 
and their
completeness is well understood. We have removed three objects (of 1486) 
because their unusually high luminosities in excess of $M_{280}=-22$ 
place them far off the main sample. These galaxies are likely to have
strong contributions from an unrecognized AGN, supported by an XMM 
detection of the one $M_{280}<-22$ galaxy which lies in the XMM area.

The galaxy sample used here covers the redshift spikes of the galaxy 
distribution in the CDFS at $z=0.67$ and $z=0.73$ \citep{Gil03,LeF04}. 
At this redshift the observed ACS F850LP band corresponds to the 
rest-frame $V$-band. Altogether, the sample contains 1483 galaxies and 
fills a comoving volume of 42375~(Mpc/$h$)$^3$. Although the slice 
selects the highest-density environments known in the CDFS, it contains 
no massive clusters. At most redshifts, the CDFS is observed to be rather 
underdense compared to the cosmic average. However, this is currently 
still the largest sample of moderately distant galaxies with redshifts 
and wide-field space-based data from which morphology can be determined. 

The photometric redshift quality of our specific sample is demonstrated
in Fig.~\ref{zz}. We have spectroscopic redshifts from the VIMOS VLT Deep 
Survey \cite[VVDS,][]{LeF04} for $\sim 10$\% of the galaxies in our sample
(with VVDS redshift reliability of 95\% or greater). We find three outliers 
among 152 galaxies (2\%) with true redshift errors on the order of 0.1,
while the remaining 98\% of the sample have a distribution of photo-z
errors $\delta_z = z_{\rm spec}-z_{\rm phot}$ of

\begin{equation}
	\delta_z/(1+z) = -0.006 \pm 0.020
\end{equation}

The COMBO-17 filters have central wavelengths ranging from observed-frame  
365 to 915~nm and allow us to calculate galaxy luminosities anywhere from
220~nm to 550~nm rest-frame wavelength directly from observed photometry
with no need for extrapolations or externally estimated $K$-corrections.
We obtain rest-frame luminosities by placing the best-fitting template SED
into the observed 17-filter spectrum and integrating the template over 
the efficiency curve of the desired redshifted rest-frame band. Here, the
template is normalised to match the total $R$-band photometry rather than
the aperture photometry to include all measurable flux from the object
and get a total luminosity. Since the shape of the SED is still determined
from aperture photometry, we measure only central colors for large, nearby
galaxies and are thus insensitive to color gradients in those. We choose
to explore the rest-frame bands Johnson $UBV$ and a synthetic 280/40-band
centered on 280~nm with a rectangular transmission function and width of
40~nm. We estimate errors for rest-frame luminosities and colors in 
COMBO-17, based on the errors of the photometry itself (for details, see 
Wolf et al. 2004). These should not exceed 0.2~mag including systematic 
aperture and calibration biases in most cases.

In Fig.~\ref{cmap} we present a completeness map for our sample, which is
used later for the calculation of luminosity functions. This map has been
constructed from end-to-end Monte-Carlo simulations of the COMBO-17 survey
which use a variety of object SEDs as an input, simulate their photometric
properties and their subsequent classification based on aperture colors as
using the standard COMBO-17 procedure. The input SEDs span the entire range 
of assumed possible spectra from the libraries of stars, galaxies and QSOs
which are used in COMBO-17 for classification and redshift estimation. The 
map shows completeness in dependence of apparent total $R$-band magnitude 
and the restframe color $U-V$. At $R_{\rm tot}>24$ our completeness is zero
because of the aperture magnitude cut ($R<24$) and because any total 
magnitude is brighter or equal to the aperture magnitude. The gentle
rise of the completeness to brighter magnitudes stems from the distribution
of aperture corrections $R_{\rm tot}-R_{\rm aper}$ and from increasing 
completeness in photo-z determination. The photo-z completeness refers to
the fact that in COMBO-17 objects are not assigned any redshift, if the
probability distribution $p(z)$ is too wide and hence the expected redshift 
error too large, which happens more often among faint objects 
(see Wolf et al. 2004 for more details). Accordingly, 
the 50\% completeness limit corresponds to $R_{\rm tot} \approx 23.7$, and 
the 90\% completeness limit to $R_{\rm tot} \approx 23$.

In terms of rest-frame selection, our observed frame cut in the $R$-band 
corresponds at $z\sim 0.7$ to a band in rest-frame centered on 380~nm and  
with 115~nm width. Apart from the width, this is similar to a rest-frame
$U$-band selection. As a last sanity check for our sample, we explore the
derived restframe colours $M_{280}-M_U$ and $U-V=M_U-M_V$, since we make 
frequent use of the $M_{280}$ and $M_V$ measurements in this paper. From
Fig.~\ref{colcol} we conclude that the entire galaxy sample follows almost 
a 1-parameter family for restframe colors between 280~nm and $V$-band.
Galaxies are either generally blue or generally red. In particular, there 
are no galaxies which combine blue $M_{280}-M_U$ colors with red $U-V$ 
colours, i.e. galaxies where the restframe $U$-band and thus the observed 
frame $R$-band would be unexpectedly faint. We conclude that average 
1-parameter color-color transformations are sufficient to describe the 
sample. Hence, it is trivial to transform our observed-frame $R$-band 
completeness map into a rest-frame $M_{280}$-band or $V$-band map.
 
\begin{figure*} 
\centering 
\includegraphics[clip,angle=270,width=0.9\hsize]{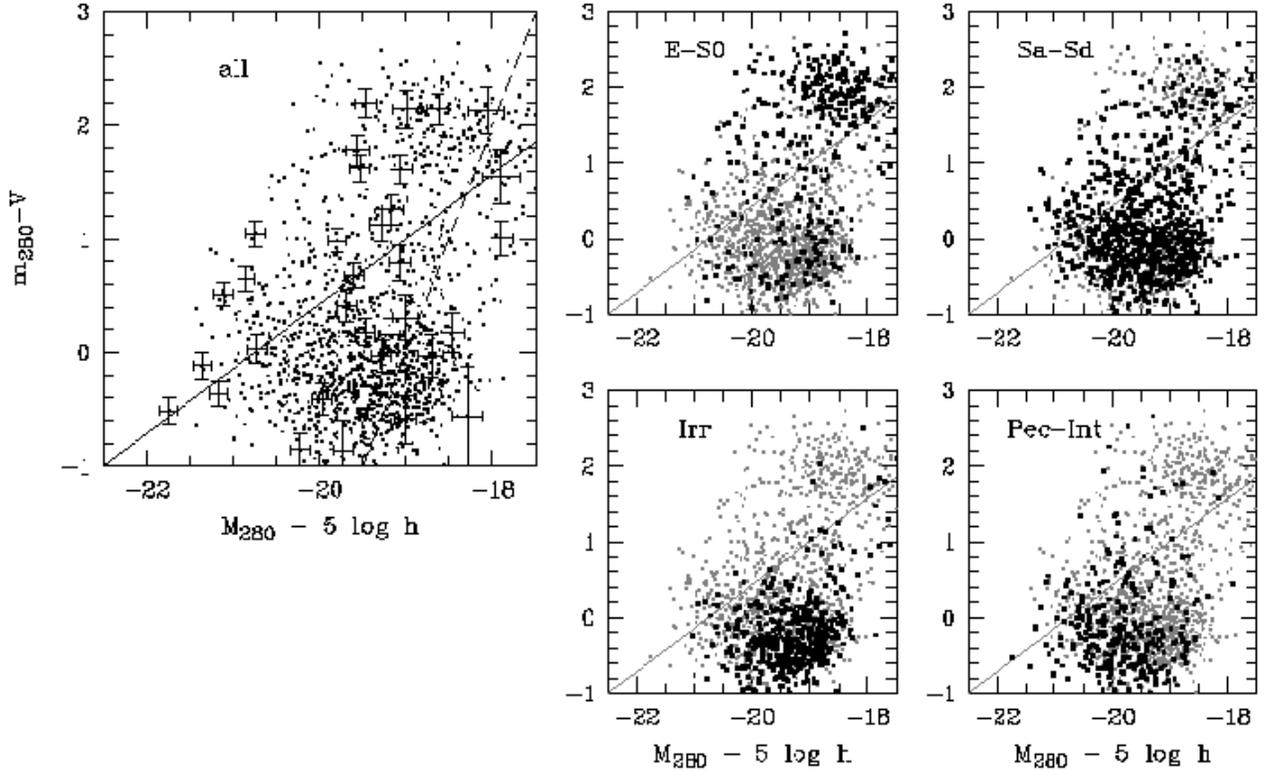} 
\caption{Color-magnitude diagram: {\it Left:} The parent sample with 1483  
galaxies.  Error bars are given for a randomly-chosen subsample; 
the error bars include expected systematic biases. 
The solid line roughly represents  
galaxies with constant mass of $10^{10} h^{-2} M_\odot$, following Bell  
\& de Jong (2001). The dashed line delineates a subsample with $>90$\% 
completeness. 
{\it Right:} Four panels showing the distribution of colors and  
magnitudes of the four different morphological types  
from all three classifiers combined (black dots) on top of the  
parent sample (grey background dots).} 
\label{cmd} 
\end{figure*}

\section{Morphological classifications} 
 
Morphological classification was carried out by eye on the F850LP images  
for the complete sample of 1483 $0.65 < z < 0.75$ galaxies. 
Galaxies were classified into a final set of four broad classes. These four  
are: {\it spheroids} (E-S0), {\it spirals} (Sa-Sdm), {\it irregulars}  
(Irr), and morphologically peculiar galaxies with disturbances suggestive  
of interaction and ongoing major mergers (Pec-Int).  
 
In Fig.\ \ref{examples} we show examples of these 
four classes.  Galaxies with dominant centrally-concentrated 
light profiles and lacking spiral structure were classified 
as elliptical (in the case where no evidence for a disk was seen) 
or S0 (in the case where a smooth disk was seen); these two classes 
are illustrated in the top panels of Fig.\ \ref{examples}. 
Galaxies with prominent disks and spiral structure 
were classified as Sa--Sd; where Sa galaxies have  
strong bulges and tightly-wound spiral structure through  
to Sd galaxies with no bulge and prominent, flocculent and  
often disorganized spiral structure (the second row of 
examples).  Galaxies with an irregular appearance suggestive 
of stochastic star formation,  
similar to Magellanic Irregular galaxies in the local 
universe, were classified as irregular (the third row of examples)
\footnote{The small number of 
galaxies which were too compact to reliably classify were  
included in the irregular bin: these galaxies were  
faint and very blue and are likely high-redshift 
analogs to the blue compact dwarf galaxies seen in the local 
universe.}.  A significant number of galaxies 
appeared to be undergoing mergers or tidal interactions: these 
were classified in the Peculiar/Interacting category (the final  
row of examples in Fig.\ \ref{examples}).  
Galaxies were classified as merging or interacting if they have 
multiple nuclei, strongly distorted inner isophotes, or large tidal 
features. These visual signatures are primarily tracers of major 
mergers (mass ratios of 1:1 to 1:4 according to Somerville et al.
2001), but also trace some minor mergers with mass ratios up to 1:10
(as assessed by doing aperture photometry of some interacting pairs). 
 
The visual classification of merging and interacting galaxies is clearly 
subjective, but automated classification presents also a serious challenge  
owing to the richness of structure in both normal spirals and interacting  
systems. Issues such as source extraction and segmentation, projection of  
very close galaxy pairs, and recognition of faint, low surface brightness  
tidal tails in otherwise largely relaxed systems are all important and not  
easily modelled automatically (see Conselice 2003 and Lotz et al. 2004 for  
current progress on some of these issues). For this reason, we chose 
to explore visually-classified morphology at this time, deferring an 
analysis based on automated classification to a later stage.  
 
We estimate the variance of this classification process by having three  
of the authors (EB, DM, CW) classify the entire sample independently.
Our fine classification scale ran along the types (E, S0, Sa, Sbc, Sd,
Irr, Pec-Int). Between authors classifications differed by $\pm 1$ class 
or less for 80\% of the sample. The classifications of galaxies in the 
faintest magnitude bin differed the most frequently, owing to the small  
size and low surface brightness of these galaxies. 
These difficulties in classifying faint galaxies led 
to two effects: first, random scatter in the classification increases as 
the by-eye classification becomes more ill-defined; secondly, at low signal 
and for small galaxies interaction features can not be discerned, which will  
bias the classification systematically against peculiar galaxies (Pec-Int). 
 
We reduce the effect of random scatter onto our results by using the 
classifications of all three authors with a weight of 1/3rd each. When 
we calculate luminosity functions, e.g., we include the author-to-author  
variation besides Poisson noise in the error analysis. This variation  
dominates errors at faint levels while Poisson noise reigns at the bright  
end, where object numbers are small and classifications agree  
most frequently. 
 
In order to assess systematic biases among faint galaxies, we also analyze 
the deeper 5-epoch GOODS images, which cover 1/5 of the entire GEMS area.  
This subsample was classified again by the same authors and compared against 
the results from the shallower 1-epoch GOODS images which are incorporated 
into GEMS. From this comparison, we have found that all classifiers find a 
significantly larger fraction of faint Pec-Int galaxies in the 5-epoch GOODS 
data than in the GEMS data. The GEMS classifications of these extra Pec-Int 
galaxies included some early types and spirals, but were dominated by  
irregulars; the extra depth afforded by GOODS allowed for the recognition of  
faint tidal features not visible in the shallower GEMS data. This source of  
uncertainty is discussed further later when determining the relative  
contributions of different morphological types to the UV luminosity density. 
In constrast, we see no significant differences among other morphological  
types, suggesting that these are easily recognized and unbiased. 
 
\begin{figure} 
\centering 
\includegraphics[clip,angle=270,width=0.8\hsize]{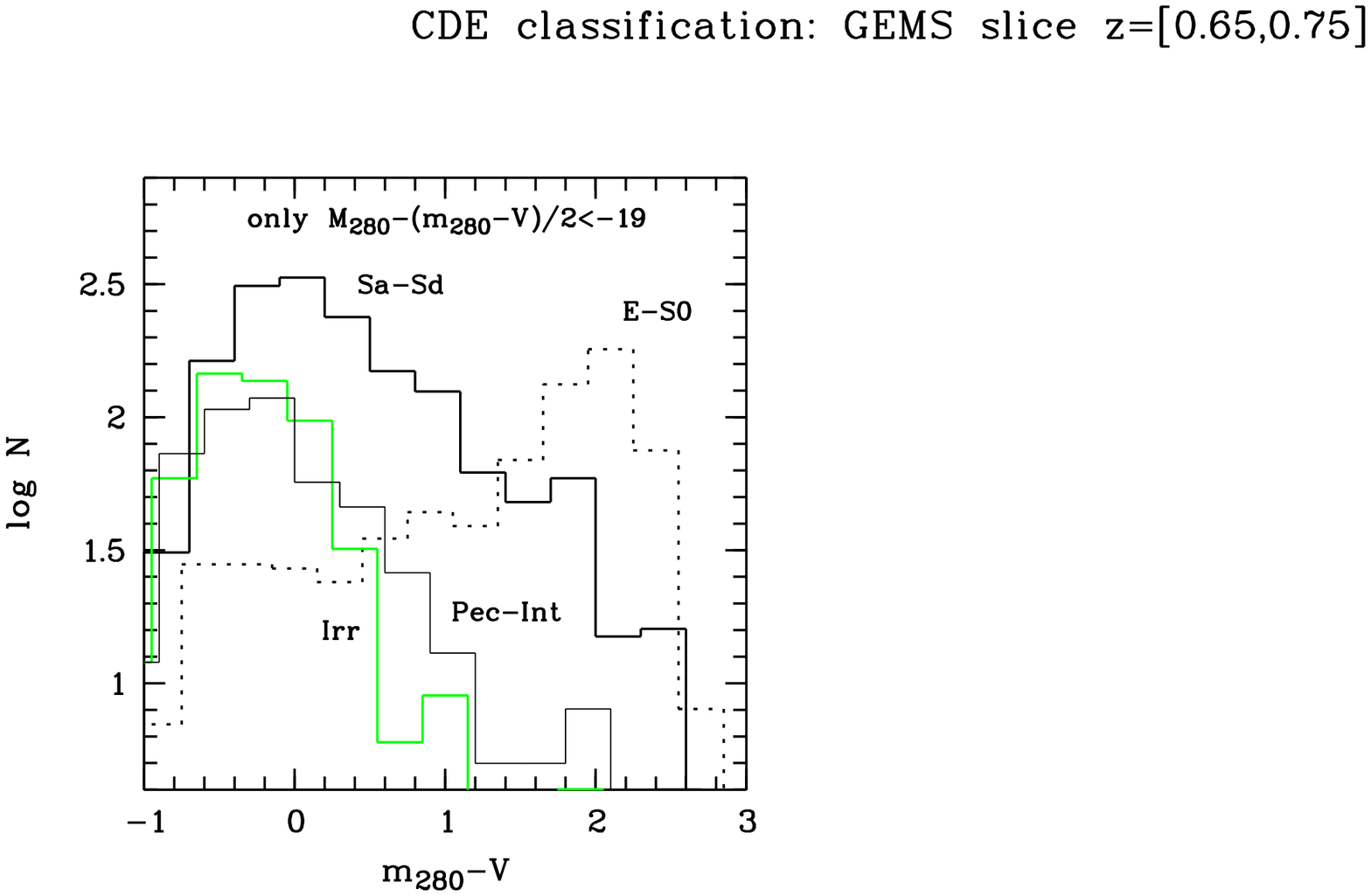} 
\caption{The rest-frame color distribution for  
different morphological types: 
Spheroids (E-S0) are predominantly red, irregulars almost uniquely very blue.  
Spirals and major mergers (Pec-Int) have somewhat similar distributions,  
with mostly blue objects and a tail to the red. Here, we only used galaxies 
from a brighter subsample that is at least 90\% complete. } 
\label{CHist} 
\end{figure}

\section{Results} 
 
\subsection{Colors and magnitudes} 
 
In Fig.~\ref{cmd} we plot the near-UV--optical color--magnitude diagrams  
(CMD) for our sample split by morphological type. The CMDs of our sample  
are bimodal, having a `red sequence' of mostly non-star-forming  
galaxies and a `blue cloud'  
of star-forming galaxies.  This bimodality of galaxy colors at optical  
wavelengths has been observed in the local Universe \citep{St01,Bal04}  
and out to $z\sim1$ \citep{Bell04}. Owing to the long wavelength base  
between 280~nm and the $V$-band ($\lambda_V/ \lambda_{280}\sim2$) and  
the sensitivity of the UV spectrum to on-going star formation, galaxies  
span four magnitudes in rest-frame $(M_{280}-M_V)$ color. 
 
The observed-frame $R<24$ selection translates into a UV luminosity 
detection limit that changes with $(M_{280}-M_V)$ color. As we can see 
in the left-hand panel of Fig.~\ref{cmd}, increasingly red  
galaxies of given visual magnitude appear at decreasing $M_{280}$  
luminosities, hence the sample looks increasingly deep towards the red.  
The sample selection has no sharp cutoff at its faint end, because the  
COMBO-17 redshift completeness declines already before reaching $R=24$. 
In Fig.~\ref{cmd} we also plot a rough cut for a brighter subsample that 
is at least 90\% complete. This particular subsample is used for the 
colour distributions plotted in Fig.~\ref{CHist}. 
 
The four right-hand panels in Fig.~\ref{cmd} demonstrate how galaxies 
of each morphological type populate the CMD. E-S0 galaxies populate 
primarily the red sequence; however, there are also some blue galaxies 
with smooth appearance and high light concentration. Examples of such 
potential blue spheroids have been seen before at a range of redshifts 
(e.g. Menanteau et al. 2001). Their fraction among the total spheroid 
population is around 30\% to 40\%, depending on restframe band ($UBV$), 
magnitude limits and choice of the red-sequence cut for their selection. 
It is interesting to note that in CMDs with optical or near-infrared 
luminosities, the most luminous galaxies are typically red 
spheroid-dominated galaxies. In the UV, these red galaxies are in fact 
somewhat fainter, on the whole, than the brightest star-forming spirals 
and interactions, owing to their color. Selecting a spheroid sample in the 
restframe UV naturally increases the fraction of blue members to, e.g., 
62\% at $M_{280}<-19$. 

Spiral galaxies (Sa-Sd) are mostly found in the blue cloud but have a 
tail extending up to the red sequence. Presumably, this tail in the 
color distribution is partly due to dust reddening, in particular for 
edge-on spirals, but in some cases reflects very low levels of ongoing 
star formation. Irregular galaxies have rather lower luminosities and 
are almost exclusively blue, presumably because of star formation and  
little overall obscuration. 
 
Visually identified mergers (Pec-Int) are concentrated towards 
the blue end of the color distribution, but  
some lie on or near the red sequence. Most interacting  
galaxies on the red-sequence show clear dust lanes, but some are  
high-inclination objects or almost regular spheroids with left-over  
tidal features. The mergers tend to have luminosities above average; 
this is partially a result of our known bias against 
faint interactions in the GEMS data (see \S~3).  However,  
a large fraction of the very bluest and most luminous  
galaxies are interactions, reflecting the important 
role of interaction-induced star formation in at least some  
fraction of galaxies. 
 
Overplotting a line of estimated constant stellar mass of $10^{10} h^{-2} 
M_\odot$ in Fig.~\ref{cmd} (simply based on observed colors and luminosities, 
following Bell \& de Jong 2001), we see irregulars predominantly below this  
cut. Spheroids are weighted towards higher stellar masses.  
Major-merger candidates spread in a range of masses around the cut.  
Disk galaxies do the same, but with a tail to low masses. 
 
Fig.~\ref{CHist} shows the number counts of the morphological samples  
against rest-frame color. Here, we use only the $>90$\% complete  
subsample to avoid any bias due to color-dependent incompleteness 
or increased classification uncertainties at faint limits. 
This selection is defined by $M_{280}-(M_{280}-M_V)/2<-19$, and is 
illustrated in the left-hand panel of Fig.~ 2 as the dashed line.  
We find the color distribution of  
Pec-Int galaxies to be somewhat similar to that of normal spirals (and  
quite different from that of irregulars). Aside from the different  
amplitude of the two distributions, the color of the Pec-Int peak is  
only slightly bluer than that of spirals and the tail to the red mimics  
that of the spirals. The similarity may reflect the compensating effects  
of enhanced star formation and enhanced dust obscuration in mergers,  
while the merging partners originate mostly from the spiral sample. 
  
\begin{figure*} 
\centering 
\includegraphics[clip,angle=270,width=0.8\hsize]{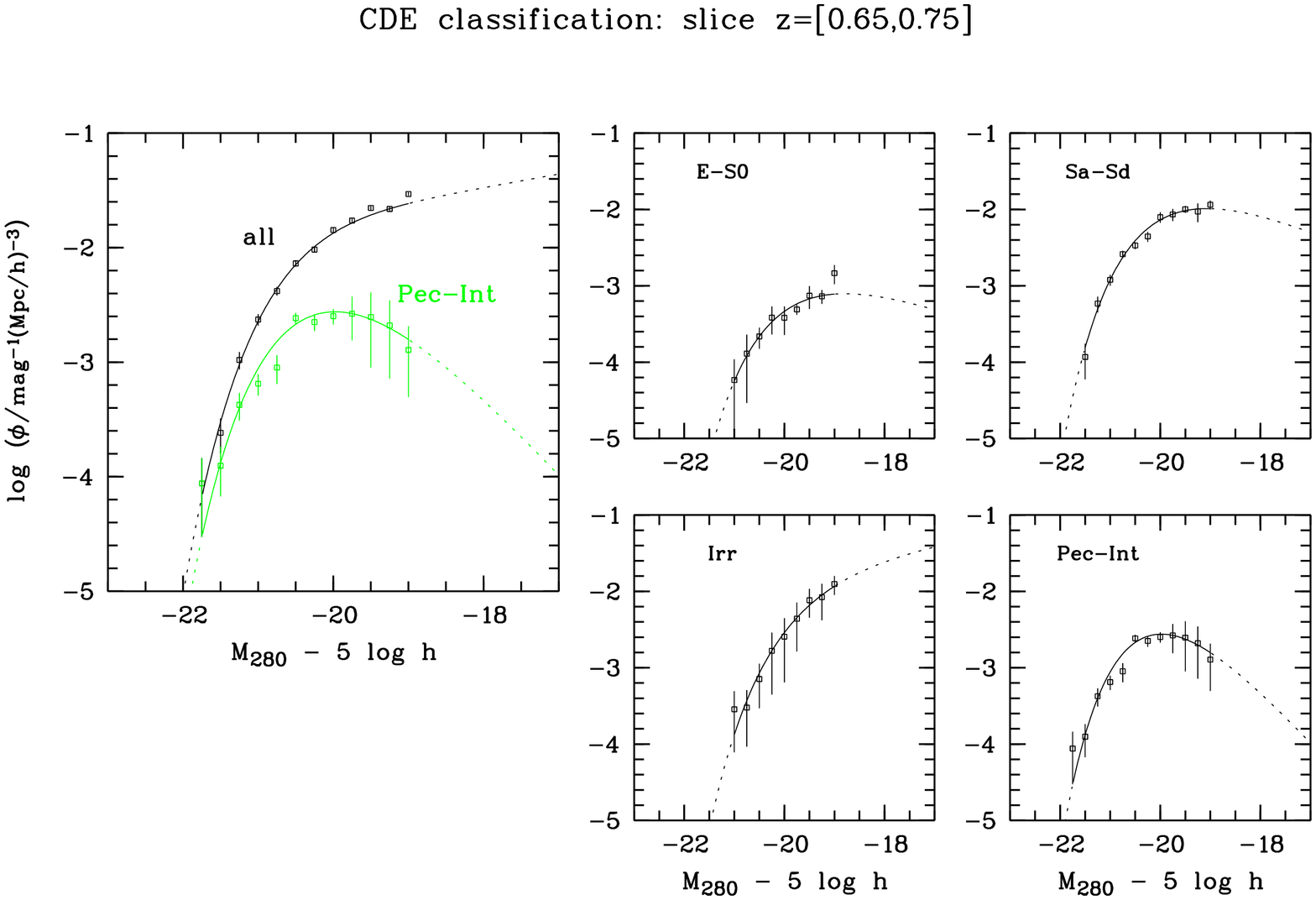} 
\caption{Near-UV luminosity functions from GEMS:  
{\it Left:} Parent sample with all galaxies, compared to clearly-interacting 
galaxies. Lines are best-fitting Schechter functions. 
{\it Right:} The four morphological types, E--S0 = spheroids, Sa--Sd =  
spiral galaxies, Irr = irregulars, Pec-Int = clearly interacting galaxies. 
Error bars are from Poisson noise or author-to-author variation from  
their independent morphological classifications, whichever is greater.} 
\label{glfs} 
\end{figure*}

\subsection{Luminosity functions} 
 
\begin{figure*} 
\centering 
\includegraphics[clip,angle=270,width=0.8\hsize]{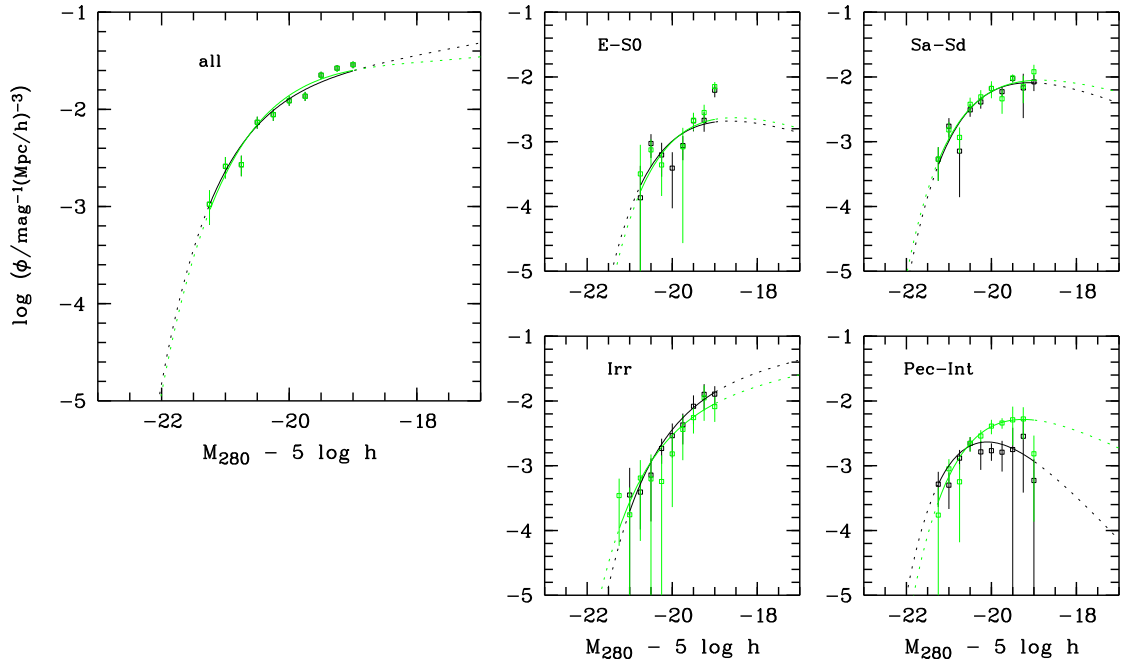} 
\caption{Near-UV luminosity functions from GOODS-shallow vs. GOODS-deep:  
The image depth influences our morphological classifications, an effect 
which propagates into the luminosity functions by type. Here, we compare 
the LFs derived from the 290 galaxies from the GOODS  
area using both GEMS-depth and GOODS-depth data  
to separate classification differences from cosmic variance. The  
GEMS result is shown in black and the GOODS result in grey. The major  
difference is that the LF of Pec-Int galaxies is increased at the faint  
end, while irregulars are lower by a similar amount.} 
 
\label{glfsGG} 
\end{figure*}

Near-UV luminosity functions (LF) for the four different galaxy types in  
the rest-frame 280~nm passband are presented in Figs.~\ref{glfs} and  
\ref{glfsGG}.
We adopt a $M_{280}\la-19$ limit for LF calculation to ensure that galaxies
of all restframe colors are represented at least with 50\% completeness.
This way we ensure that the correction applied to $\phi(M)$ from our 
completeness maps never exceeds a factor of two. This cut corresponds
to $R\la 23.5$ for the bluest star-forming galaxies and to $R\la 22$ for
red-sequence galaxies (see Wolf et al. 2003 for all equations and  
technical details on completeness and luminosity functions).

We present three sets of rest-frame UV LFs in this paper. In Fig.~\ref{glfs}, 
we show the LFs of the entire GEMS sample, split by morphological type.   
However, comparison between classifications derived from the $5\times$  
deeper GOODS data and GEMS classifications demonstrated that the GEMS data  
are less sensitive to lower-luminosity interactions than the GOODS data.   
Unfortunately, the GOODS area is only 1/5 of the GEMS area, resulting in  
increased uncertainties from large-scale structure. We therefore address  
the classification bias, independent of large scale structure uncertainties,  
by presenting a differential deep-vs.-shallow comparison of GOODS- and  
GEMS-derived LFs for only the subsample of 290 galaxies in the GOODS area. 
Altogether, we present LFs for three different samples: 
 
\begin{itemize} 
\item  {\it GEMS:} the full area with 1483 galaxies classified on shallow  
    images, providing small statistical errors as shown in Fig.~\ref{glfs} 
\item  {\it GOODS-shallow:} the central sub-area of the GOODS field with 
    290 galaxies, classified on shallow GEMS-quality images 
\item  {\it GOODS-deep:} the GOODS field with 290 galaxies, classified on 
    deep GOODS images. 
\end{itemize} 
The LFs of GOODS-shallow and GOODS-deep are compared in Fig.~\ref{glfsGG} 
to assess the impact of classification bias when shallow images are used 
for galaxy classification.  
 
For each LF we plot $V_{max}$ values with error bars derived from Poisson  
statistics or from the author-to-author variation of the space density,  
whichever is greater. We overplot the best-fitting Schechter  
functions from an STY-style fit to the sample \citep{STY} and list their 
parameters in Table~\ref{STYpars}.  In the left-hand panel of 
Fig.~\ref{glfs} where the LF for all galaxies is shown, the 
lines denote the sum of the four individual type-split LFs. 
 
Ideally, we would like to derive the Schechter function  
slopes independently for each  
data set. However, for some purposes  
the GEMS observations lack the sufficient depth  
and for some the GOODS area is not  
wide enough to constrain the LF shapes with high accuracy.  
Thus, we combined findings from both 
data sets to constrain the final set of Schechter functions. Hence we discuss  
now both Fig.~\ref{glfs} and Fig.~\ref{glfsGG} together and list all the 
fine-tuning steps we have applied in STY fitting here: 
 
{\bf E/S0:}  
The GOODS-deep sample is too small to get any useful STY fit, while the GEMS  
sample is not deep enough. Both yield unrealistic slopes 
of $\alpha\sim-2$. Hence, we adopt the slope measured by Cross et al.
(2004) from their deeper sample, which is $\alpha = 0.75 \pm 0.13$ for their
combined sample of red and blue spheroids. By constraining our STY fit to
this slope, we determined $M^*$, $\phi^*$ and $j_{280}$ from our data.
 
{\bf Sa--Sd:} 
The Schechter function for spirals is well-constrained at least in GEMS  
and plausible in both samples ($\alpha_{\rm GEMS}=-0.32\pm 0.17$ and 
$\alpha_{\rm GOODS-deep}=-0.52\pm 0.41$). The LFs are determined  
independently. 
 
{\bf Irr:} 
In the local Universe, irregular galaxies have very steep LFs with faint characteristic luminosities, making it almost impossible to estimate robust 
Schechter parameters if the sample is not both deep and large. 
Neither the GEMS nor the GOODS sample yielded particularly 
well-constrained faint-end slopes: 
We find $\alpha =-0.92 \pm 0.35$ from GEMS and $\alpha =-1.83 \pm 0.40$ 
from GOODS-deep. Since the $V_{max}$ data points do not differ much and do  
not suggest a completely different LF, we ended up using the weighted mean  
of the two values, i.e. $\alpha_{\rm Irr}=-1.31\pm 0.20$, for all Irr LFs. 
 
{\bf Pec-Int:} 
The slope of the Pec-Int LF in GEMS is $\alpha =+0.82\pm 0.34$, which is 
biased by the incompleteness of GEMS in picking up faint mergers.  
For the GOODS-deep STY fit, the last data point is discarded owing to its 
large uncertainties, yielding a faint-end slope $\alpha = -0.15 \pm 0.50$. 
The increase of Pec-Int galaxies in GOODS-deep requires a corresponding  
decrease in other classes, because total galaxy number is conserved. The  
decrease is most pronounced among the irregular galaxies. 
 
Altogether, it appears that visual classification of GEMS images biases 
one against faint Pec-Int galaxies. Other galaxy types are easily recognized  
and may be randomly confused at very low S/N, but are probably unbiased.  
The slopes of their Schechter functions are broadly consistent with the  
vast literature body of local values for blue bands.  
 
Comparing the GOODS-shallow LF from Fig.~\ref{glfsGG} with the GEMS LF in 
Fig.~\ref{glfs} demonstrates the effect of cosmic variance in such small  
fields. In the small GOODS field we see a 3-fold increase in the space  
density of spheroids, which probably reflects the presence of several  
known galaxy concentrations at $z\sim 0.7$ in the central GOODS field. In 
contrast, we find slightly decreased abundances of spirals and Pec-Ints.

\begin{table*} 
\begin{center} 
{\begin{tabular}{lr@{$\pm$}rr@{$\pm$}rr@{$\pm$}rr@{$\pm$}r@{}rc} 
\noalign{\smallskip} \hline \noalign{\smallskip}\hline \noalign{\smallskip} 
sample  & \multicolumn{2}{c}{$M^*-5$~log~$h$} &  
  \multicolumn{2}{c}{$\phi^* \times 10^{-4}$} &  
  \multicolumn{2}{c}{$\alpha$} &  
  \multicolumn{3}{c}{$j \times 10^{7} L_{\odot}$} & $c_{\phi^*,L^*}$ \\ 
 & \multicolumn{2}{c}{(Vega mag)} & \multicolumn{2}{c}{$(h/$Mpc)$^{-3}$} & 
  \multicolumn{2}{c}{} & \multicolumn{3}{c}{$(h/$Mpc$^3$)}  \\ 
\noalign{\smallskip} \hline \noalign{\smallskip} 
GEMS \\ 
\noalign{\smallskip} \hline \noalign{\smallskip} 
E-S0    & $-19.53$ & $ 0.19$ & $   22.83$ & $   6.56$ & $-0.75$ & $ 0.13$ & $  6.22$ & $ 1.79$ & $_{-0.22}^{+0.29}$ & $ -0.906$ \\ 
Sa-Sd   & $-19.49$ & $ 0.11$ & $  285.93$ & $  45.45$ & $-0.32$ & $ 0.17$ & $ 74.28$ & $11.81$ & $_{-1.74}^{+2.43}$ & $ -0.779$ \\ 
Irr     & $-19.33$ & $ 0.17$ & $  239.70$ & $ 132.52$ & $-1.31$ & $ 0.20$ & $ 78.53$ & $43.41$ & $_{-14.19}^{+23.95}$ & $ -1.686$ \\ 
Pec-Int & $-19.33$ & $ 0.16$ & $   62.04$ & $  19.17$ & $ 0.82$ & $ 0.43$ & $ 26.20$ & $ 8.10$ & $_{-0.81}^{+1.40}$ & $ -0.162$ \\ 
all     & $-19.67$ & $ 0.16$ & $  556.35$ & $  51.87$ & $-0.75$ & $ 0.20$ &  
% $171.65$ & $16.00$ from Schechter fit  
$185.23$ & $45.73$ & $_{-14.33}^{+24.13}$ & $ -0.943$ \\  % the sum of four 
\noalign{\smallskip} \hline \noalign{\smallskip} 
GOODS \\ 
\noalign{\smallskip} \hline \noalign{\smallskip} 
E-S0    & $-19.25$ & $ 0.37$ & $   93.35$ & $  11.74$ & $-0.75$ & $ 0.13$ & $ 19.56$ & $ 2.46$ & $_{-0.90}^{+1.12}$ & $ -1.140$ \\ 
Sa-Sd   & $-19.64$ & $ 0.31$ & $  222.27$ & $  61.72$ & $-0.52$ & $ 0.41$ & $ 64.89$ & $18.02$ & $_{-3.63}^{+8.64}$ & $ -0.799$ \\ 
Irr     & $-19.65$ & $ 0.36$ & $  141.19$ & $  89.27$ & $-1.31$ & $ 0.20$ & $ 61.86$ & $39.11$ & $_{-9.18}^{+16.50}$ & $ -1.417$ \\ 
Pec-Int & $-19.43$ & $ 0.37$ & $  151.17$ & $  48.52$ & $-0.15$ & $ 0.50$ & $ 39.01$ & $12.52$ & $_{-2.15}^{+4.15}$ & $ -0.717$ \\ 
all     & $-19.64$ & $ 0.26$ & $  558.63$ & $  70.24$ & $-0.75$ & $ 0.20$ &  
% $166.89$ & $20.98$ from Schechter fit 
$185.32$ & $44.95$  & $_{-10.20}^{+19.29}$ & $ -0.975$ \\  % the sum of four 
\noalign{\smallskip} \hline 
\end{tabular}}
\caption{Results of STY fits of the 280~nm luminosity functions: Listed  
are $M^*$, $\phi^*$ with its author-to-author variation, $j$ and the  
covariance between $\phi^*$ and $L^*$. The $j$ values for the combined  
sample are sums over the $j_{\rm type}$. All $j$ values are given  
in units of the solar luminosity within the 280/40 passband, which is $M_{ 
280}=6.66$ in Vega units or $2.56\times 10^{10}$~W/Hz. A first set of 
errors in $j$ is propagated from errors in $\phi^*$; the set of smaller, 
asymmetric errors reflect the propagation of $\alpha$ errors into $j$. 
\label{STYpars}} 
\end{center} 
\end{table*}

\subsection{UV luminosity density} 
 
\begin{figure*} 
\centering 
\includegraphics[clip,angle=270,width=0.8\hsize]{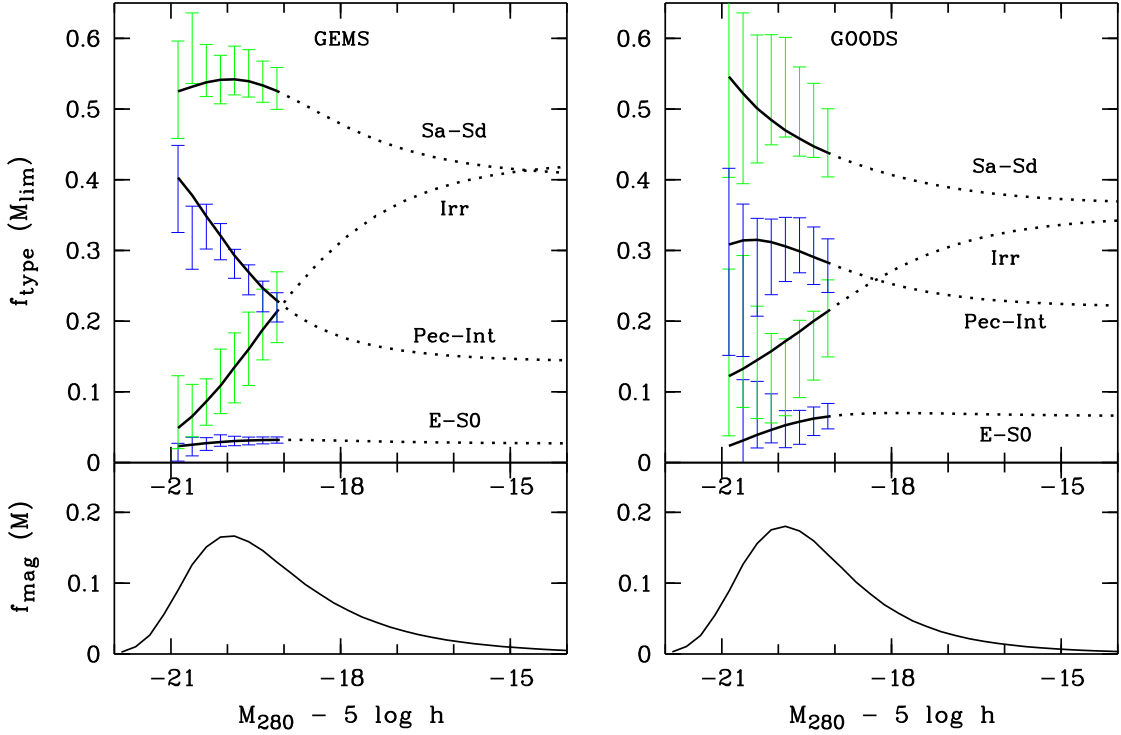} 
\caption{Fractional contributions to the 280~nm luminosity density from 
different morphological types, as derived from the wide-area GEMS mosaic 
(left) and the deeper, but smaller GOODS data (right): 
{\it Top panels:} Contribution $f_{\rm type} (M_{\rm lim})$ of the four  
galaxy types to $j_{280}(M_{\rm lim})$, depending on the depth to which 
one integrates. Thick lines are derived from the Schechter function fits.
Error bars are centered on the LF data themselves and show the greater of 
either Poisson noise or author-to-author variation in the morphological
classification. Two thirds of the total extrapolated light is directly 
probed by the galaxy sample. At $M_{280}\sim -15$ the total luminosity 
density $j_{280}$ and the type-dependent fractions have converged. 
{\it Bottom panels:} Contribution $f_{\rm mag} (M)$ of magnitude intervals  
to the total integrated 280~nm luminosity density which peaks at $M_{280} 
\sim -20$, while most UV light comes from galaxies at $-21<M_{280}<-18$.} 
\label{j280} 
\end{figure*}

Our data can directly address which galaxies are responsible for producing  
most of the UV luminosity at redshift $z\sim0.7$, depending on  
their luminosity and their morphological type. We first determine how the 
contribution to the integrated UV light depends on galaxy luminosity using 
the Schechter fits across a wide range of magnitudes. Here, we define 
 
\begin{equation} 
  f_{\rm mag}(M) = \frac{\int_{M-0.25}^{M+0.25} L(M) \phi(M) dM}   
			{\int_{-\infty}^{+\infty} L(M) \phi(M) dM}  ~. 
\end{equation}  
The bottom panels of Fig.~\ref{j280} show this fraction $f_{\rm mag}$ as  
derived from the LFs of GEMS and GOODS samples; as the total GEMS and 
GOODS LFs are very similar in shape, the $f_{\rm mag}$ distributions are 
almost indistinguishable. Galaxies with $M_{280}\sim-20$ make the 
strongest contribution to $j_{280}$; the bulk of the UV light is emitted 
in the luminosity range $[-21,-18]$. Our observed galaxy sample itself 
provides 66\% of the total UV luminosity density integrated from 
$-\infty$ to $+\infty$, which is $(4.74\pm 1.17) \times 10^{17}$~W/Hz. 
 
We can now estimate the fraction of UV luminosity density contributed 
by different galaxy types. These numbers depend on the depth to which we 
integrate, owing to the different Schechter parameters of the different 
galaxy types. The luminosity densities and their fractions by type are 
here defined in dependence of the limiting depth $M_{\rm lim}$ by 
 
\begin{equation} 
  j_{\rm type}(M_{\rm lim}) =  
	{\int_{-\infty}^{M_{\rm lim}} L(M) \phi_{\rm type}(M) dM} 
\end{equation}  
and 
\begin{equation} 
  f_{\rm type}(M_{\rm lim}) = \frac{j_{\rm type} (M_{\rm lim})} 
	{\sum_{\rm type} j_{\rm type} (M_{\rm lim})}  ~. 
\end{equation}  
The top panels in Fig.~\ref{j280} show the fractions $f_{\rm type}$  
derived from the Schechter fits to the luminosity function as lines;  
solid lines for the measured part and dotted lines where the LF  
is extrapolated.  Because of the relatively shallow faint-end slope 
of the type-split LFs, $f_{\rm type}$ converges reasonably quickly. 
 
Error bars show the 1-$\sigma$ confidence intervals of $f_{\rm type}$ as 
derived from the $\phi(M)$ measurements for the full GEMS data set (left)  
and for the deeper, smaller GOODS data (right). The error bars are purely 
statistical, reflecting either the Poissonian limitations of the sample  
size or the author-to-author variation in the classification, whichever  
is greater.  They do not include contributions from cosmic variance or  
any systematic biases in classifications due to image depth. 
 
With respect to these two error sources, the data from GEMS and GOODS  
play largely complementary roles, with each one having one advantage  
over the other: GEMS has a $5 \times$ larger field, reducing the impact  
of cosmic variance, whereas GOODS has substantially deeper imaging data,  
yielding lower classification uncertainties at faint limits. 
 
Our sample contains 66\% of the total luminosity density $j_{280}$ of  
the Universe at $z\sim 0.7$. By integrating $j_{280}$ only across the  
complete sample of galaxies in the GEMS sample at $M_{280}<-19$,  
we find more than $52\pm 3$\% of the light originating from spirals 
(author-to-author variation including Poisson noise). Irregulars and 
Pec-Int galaxies contribute $\sim 22$\% each ($\pm 5$\% and  
$\pm 2$, respectively), while $3.5\pm 0.5$\% is from spheroids. 
 
The GOODS sample has statistically larger error bars on all types, but  
less bias against faint Pec-Ints.  
Consequently, the GOODS sample yields an increased contribution  
by Pec-Ints to the luminosity density of $26\pm 4$\%.  
This is balanced by a decreased spiral contribution of $44\pm 5$\%; 
spheroids contribute $9\pm 2$\%.  
The fraction attributed to irregulars is largely unchanged at $21 
\pm 5$\%. Where these trends do not reflect those expected from what 
we expected to be primarily a re-labeling of some fraction of  
faint irregulars into mergers, they simply reflect the 
extent by which the GOODS area deviates from the cosmic average. 
 
Fig.~\ref{j280} clearly shows that the relative contributions of 
different galaxy types is relatively insensitive to extrapolation 
of the UV LFs to $M_{\rm lim} \rightarrow \infty$. Spiral galaxies 
are still an important 
source of UV light, accounting for $\sim 40$\% of $j_{280}$, whereas 
E-S0 galaxies account for $\la 5$\%.  Owing to the steepness 
of the faint-end slope of irregular galaxies, their contribution 
to $j_{280}$ may be as high as $\sim 40$\%.  In contrast, owing 
to the decreasing importance of clear interactions towards fainter 
limits, even in the GOODS data, less than 25\% of $j_{280}$ is from 
clearly interacting galaxies, with a value of $\sim 15$\% in 
the GEMS sample alone. 
 
The propagation of errors from the LF parameters into the luminosity 
density is listed in Tab.~1, but not plotted in Fig.~\ref{j280}. Two
sources of error are explicitly addressed, which are the uncertainty in
$\alpha$ and the error in $\phi^*$ at fixed $M^*$, which reflects just
the author-to-author variation or Poisson errors whichever is greater. 
The error in $M^*$ is not propagated into $j$ explicitly, because it is 
directly coupled with a compensating change in $\phi^*$. The two effects
cancel, if the covariance $c_{\phi^*,L^*}=-1$, which is roughly the case.
The error in $\alpha$ is propagated by refitting the LF with $\alpha^\pm$
set to $\alpha \pm \sigma_\alpha$. In this procedure $M^*$ will change 
due to the coupling with $\alpha$, and $\phi*$ will change to accommodate
the new shape of the fit to the measured data points. As a result, the
luminosity density will not change significantly in the measured range of
luminosities, but only in the extrapolation range. The largest uncertainty
comes from the $\phi^*$ error itself which propagates linearly into $j$.

We do not actually know the shape of the LF of really faint galaxies at 
$z=0.7$, but we can play with an unlikely and extreme thought experiment. 
Presume that {\it all} the 
faint galaxies that contribute the unobserved one third of the total 
UV light belong to one single morphological class, whereas the other 
three respective classes exist only at the observed bright magnitudes. 
While this scenario is completely unrealistic, it would increase the  
$j_{280}$ fraction of the then dominant type above the previously stated  
numbers by the highest possible degree. Rotating through the four classes  
as sole constituents of faint galaxies, we find their absolute-maximum 
values $f_{\rm type, max} \approx (35\%,65\%,50\%,50\%)$ for (E-S0, Sa-Sd, 
Irr, Pec-Int), respectively.

\section{Discussion} 
 
We motivated our analysis with the observation that at the present day 
the 280~nm luminosity density is a factor of 3--6 lower than at $z=0.7$.  
This drop must reflect a strong decrease in the UV luminosity of the  
largest contributors at $z=0.7$, i.e. the spirals and the faint blue 
irregulars. A drop in the UV luminosity of ongoing mergers is expected  
also; however, they account for less than a quarter of the UV light at 
$z\sim 0.7$ to start with. Therefore, the evolution of such mergers 
(even if they had completely disappeared by $z=0$) can not be the primary 
driver of the overall change in the UV luminosity density and the 
unobscured SFR of the galaxy population.

\subsection{Comparison of galaxy colors}

The fraction of spheroids with blue colors has been measured by various
authors at around 30\% to 40\%, and is possibly not a strong function of 
redshift between $z\sim 0.3$ and $z\sim 1$ (Cross et al. 2004; Menanteau 
et al. 2004; Stanford et al. 2004). We measure fractions between 30\% 
to 40\% as well, depending mostly on the choice of restframe band ($UBV$) 
and magnitude limits but also on the precise choice of the red-sequence 
cut. Stanford et al. (2004) found more than 50\% blue members in a 
spheroid sample selected in the NIR on NICMOS images. However, many of 
their blue spheroids appear like late-type galaxies in optical bands from 
WFPC2. Menanteau et al. (2004) investigated spatially resolved color maps 
of spheroids at $z\la 1$ from HST images and found predominantly blue 
cores but also inhomogeneous internal colors suggestive of an early stage 
in the formation of spheroids.

We found the merger candidates to have a color distribution similar to that 
of spirals with only a slight shift to the blue, while the irregular galaxies
are distinctively bluer than both the average merger and average spiral.
The same result has been reported by Conselice et al. (2003). This
distribution has also been found in the local Universe by Bergvall et al.
(2003), who found the colors of interacting galaxies to be only slightly
bluer than those of non-interacting galaxies at $z\sim 0$. They observed 
also similar $L_{\rm FIR}/L_B$ ratios in both samples, leaving no room
for strong obscured starbursts.

\subsection{Comparison of luminosity densities} 
 
The HST observations in the Canada-France Redshift Survey (CFRS) fields  
were seminal in this field, but limited by sample size and depth. In 
their first paper, Brinchmann et al. (CFRS1, 1998) examined  
the types of galaxies contributing to the $B$-band luminosity from  
$z\sim 0.9$ to $z\sim 0.3$. CFRS1 used three galaxy types, i.e. {\it  
spirals}, {\it ellipticals} and {\it irregulars} including ongoing  
violent mergers, and found most of the decline in $j_B$ to be a result of  
the evolution of spiral galaxies, which dominate $j_B$ at the high-z end.  
Brinchmann et al. found that the $B$-band contribution from irregulars  
dropped at least as much as that of the spirals, although they provided 
a smaller contribution to begin with. 

While this is fully consistent with our observations, the $B$-band  
luminosity reflects a broad range of stellar ages and the decline in
luminosity density is less pronounced than in the UV. Dickinson et al. (2003)
found that $j_B$ declines by a factor of 2.5 from $z\sim 1$ to $z\sim 0$ and
Wolf et al. (2003) found a factor of $\sim 2$, which are flatter 
than the decline in the UV. Our results, determined in 
the UV, show that irregulars plus visible mergers contribute more to 
$j_{280}$ then the spirals, once we include the faint regime.  
This is a natural result of irregulars being bluer on average than  
spirals and appearing relatively more luminous in the UV. 

In terms of stellar mass, Brinchmann \& Ellis (2000) found roughly $1/3$
of the stellar mass to be in each of spheroids, disks and irregulars at
$z\sim 1$. Towards the present day the mass fraction in irregulars drops
to $\sim 2$\% of the total. The disappering stellar mass from irregulars
appers as a mass increase among the regular morphological types, which is
either a result of merging or a consequence of more regular star formation
patterns being established in an initially irregular galaxy.

In a recent paper, Conselice, Blackburne and Papovich (2005) finally show
that 90\% of the blue luminosity density at $z\sim 1$ originates from 
spirals and spheroids in roughly equal parts, leaving only 10\% to 
irregulars and merging galaxies. They conclude that at $z<1$ galaxies
grow mostly through minor mergers or quiescent star formation driven by 
gas infall, while major mergers are only dominating at $z>2$.

\subsection{Comparison of merger fractions} 
 
The fraction and rate of galaxy mergers was estimated by various authors 
at different redshifts and luminosities. Le F\`evre et al. (CFRS4, 2000) 
used visual classification of the CFRS galaxy sample observed with HST. 
They found a dramatic increase in galaxies with merger morphologies 
towards higher redshift, reaching a fraction of 10\% at $z\sim0.63$ and 
even $\sim20$\% at $z\sim0.9$ (selected with $I_{AB}<22.5$). Their 
results reflect the merger rate at the high-luminosity end, where we have 
also found interacting galaxies to be quite abundant. If we limit our 
sample to $I_{AB}=22.5$, then we find a merger fraction of 14\% at 
$z\sim0.7$, consistent with those of CFRS4. We note that at $z\sim 
0.7$ the observed-frame $I$-band selection corresponds to a rest-frame 
$B$-band selection, which is also used by several other studies.

CFRS4 conjectured that mergers could be even more abundant among galaxies 
of lower luminosity. Our analysis implies that this is not the case: 
Although recognizable mergers are very common at the highest of all 
UV galaxy luminosities, their LF rises less quickly than the LF of  
the total galaxy population. Therefore, the integrated fraction of  
classifiable mergers actually drops with fainter UV luminosity.  

While incompleteness of mergers among faint galaxies is a concern,
contamination due to projection effects or large star-forming regions 
may also be a problem. In fact, Bundy et al. (2004) suggest that major
merger and close pair fractions determined in the optical at $z\sim 1$ 
may be inflated by bright star-forming regions that are unlikely to be 
representative of the underlying mass distribution. Using $K$-band
observations they derive the fraction of pairs that will produce major
mergers within $\sim 1$ Gyr. These fractions are much lower than optical 
bands would suggest for the same galaxy sample, because many companions
in the optical are particularly blue. Hence, they have lower masses than
suggested by their light contribution, and lead only to minor mergers.
As a result of selecting in the $K$-band, they find less than 10\% mergers
at $z\sim 1$, where HST $I$-band data suggested $\sim20$\%.

Conselice et al. (2003) measured the merger rate using automated measures of
morphological disturbance in the rest-frame $B$-band from $z \sim 3$ to the 
present day using the Hubble Deep Field North (Williams et al. 1996).  
Their results show an increasing merger fraction with increasing luminosity; 
at $z \sim 0.6$ they find a 4\% merger fraction at $M_B < -18$ and 
7\% mergers at $M_B < -20$.  Their overall fraction is lower than the 
one derived in this paper, but bearing in mind the small field size
and differences in methodology, the agreement is not unreasonable.

Studies of close pairs of galaxies can also constrain merger rate.
To date, they have typically inferred rather lower
merger rates than morphological studies.  Patton et al. (2002) 
found that 15\% of galaxies with $-21 < M_B < -18$ have undergone
a major merger since $z \sim 1$ using dynamically-close pairs
from the CNOC2 survey.  Recently, Lin et al. (2004) used a sample
of wide-separation pairs to estimate the merger rate, finding that 
rather fewer than 10\% of present-day $L^*$ galaxies have undergone 
a major merger since $z \sim 1$, although with considerable 
uncertainties owing to the use of wide-separation pairs.  They
suggested that studies using morphological criteria to identify 
mergers may pick up many minor 
mergers as well, which should help to explain the disagreement. 

Irrespective of whether major mergers are defined morphologically or via 
close pairs, it appears that recent work converges to the conclusion that
most stellar mass formed at $z<1$ is not formed via starbursts in major 
mergers, while the opposite may well be true at $z>2$ \cite{Con03b,Con05}.

\subsection{Effects of dust extinction} 
 
The impact of extinction by dust is a significant uncertainty  
affecting the interpretation of rest-frame UV luminosities. 
This is of particular concern in 
this study, as dust extinction is known to depend on gas content, 
metallicity, geometry, and stellar age --- all factors that correlate 
strongly with morphological type. Therefore, we can not immediately 
conclude that contributions of various galaxy types to the UV 
luminosity density reflect their relative contributions to the  
cosmic-average SFR.  
 
Yet, an understanding of the contribution of different galaxy 
types to the UV luminosity density is important, both because 
it is possible to reach down well below the knee of the luminosity 
function, and because an understanding of the UV luminosity density 
gives some insight into, e.g., plausible sources of ionizing  
UV photons.  Bell et al. (2005) use deep 24$\micron$ data  
from Spitzer to explore the importance of obscured star formation  
for the same $0.65<z<0.75$ sample of galaxies used in this paper.  
Roughly 1/3 of the GEMS sample are detected at 24{\micron}, accounting 
for more than half (likely $\sim$2/3) of the total SFR at $z \sim 0.7$. 
They find many highly-obscured galaxies, with IR/UV $\ga 10$. 
Yet, despite this dominant contribution from the IR, the 
qualitative picture is very similar: $\la 40$\%, and more likely 
$\sim 30$\%, of the total SFR (as derived from the IR and UV luminosities 
combined) is in clearly-interacting systems, with the rest 
coming primarily from spiral galaxies with small contributions 
from irregulars and E/S0 galaxies.  Their results are fully 
consistent with earlier ISO results (e.g., Flores et al.\ 1999; 
Zheng et al.\ 2004), with the important advantages of increased 
sample size and sensitivity. This supports the primary conclusion of 
this paper: a decreasing major merger rate cannot be the dominant driver 
of the drop in cosmic SFR between $z\sim 1$ and the present day.

\subsection{Expectations from hierarchical models}

The hierarchical structure formation paradigm within a given CDM
cosmology makes specific predictions for the evolution of the merger
rate of \emph{dark matter halos} as a function of time. However,
converting this to quantities that can be directly compared with the
observations presented here involves several uncertain steps. First,
the amount of star formation associated with mergers depends on the
details of the star formation and feedback recipes implemented in the
models. Second, we need to distinguish between two questions, one of
which is physically more interesting and the other of which is what is
observationally addressed in this paper:

\begin{enumerate}

\item What fraction of the total star formation at a given epoch is
  caused by a merger-triggered burst?

\item What fraction of the total star formation is contributed by
  galaxies that appear morphologically disturbed?

\end{enumerate}

Based on hydrodynamic simulations, it is likely that the timescale
over which galaxies appear morphologically disturbed following a
merger is considerably longer (about a factor of a few to ten) than
the timescale over which star formation is significantly enhanced
(e.g. Cox et al. 2005). As well, galaxies with low gas fractions may
not experience a significant star formation episode in a merger event.

In order to address this question in a way that can be directly
related to the observational results presented here, we made use of a
GEMS mock catalog created using a semi-analytic model similar those
presented in Somerville, Primack, \& Faber (2001, hereafter SPF). The
global star formation rate density produced by this model is in good
agreement with dust-corrected observational estimates from $z\sim0$--6
(SPF; Giavalisco et al. 2004), and the model also reproduces many
(though not all) observational properties of low and high redshift
galaxies. We assume that a galaxy would be identified as
``morphologically disturbed'' if it has experienced a major merger
(1:4 or greater mass ratio) within a timescale $\tau_{\rm dist} = \eta
t_{\rm dyn}$, where $t_{\rm dyn}$ is the internal dynamical time of
the galaxy and $\eta$ is a factor of order a few. Here we use
$\eta=2$. We consider galaxies in a redshift slice $0.65<z<0.75$, as
for the GEMS sample.  For a model similar to the ``collisional
starburst'' model described in SPF, in which both major and minor
mergers trigger bursts of star formation, we find that 30 percent of
the star formation is contributed by these ``strongly disturbed''
galaxies. We find that 53 percent of the star formation is contributed
by galaxies that have experienced a merger down to a mass ratio of
1:20 within the timescale $\tau_{\rm dist}$ systems \footnote{It is
  worth noting, however, that even in a model in which we assume that
  minor mergers do not trigger bursts, we find that 51\% of the star
  formation is ``associated'' with minor mergers. This illustrates the
  ever present danger of equating correlation with
  casuality.}. Varying the recipes for quiescent and burst star
formation to encompass alternative choices commonly made in the
literature (see e.g. SPF), we obtain a range of values between 12 and
33 percent for the fraction of star formation in strongly disturbed
galaxies (major mergers). This result does not change significantly
for values of $\eta$ in the range 1--10, and the fractions are similar
for a magnitude limited sample selected with the GEMS/COMBO limit
$R<24$. We present a more detailed description of the models and the
dependence on the model ingredients in a subsequent paper (Somerville
et al. in prep).

We can compare these results with the predictions that may be read
directly from Fig.~10 and 11 of SPF, which show the star formation
rate contributed by quiescent and actively bursting galaxies. Based on
those figures, the collisional starburst model predicts that 35 \% of
star formation at $z=0$ is due to bursts in galaxy interactions of all
mass ratios, while only 2.5 \% arises from major (4:1 or greater)
mergers. At $z=0.7$, about 50 \% of the star formation is from bursts
of all types, with only 5\% arising from bursts associated with major
mergers. At $z=3$, the model predicts that 75 \% of all star formation
is due to collisional starbursts, while 18 \% is due to major-merger
induced bursts. It is important to keep in mind, however, that the
timescale of the starburst assumed in this tabulation was assumed to
be $\sim 0.1$--1 $t_{\rm dyn}$. This factor of $\sim 2--10$ difference
between the duration of the burst and the duration of the
morphologically disturbed phase, assumed above, accounts for most of
the difference between the values of 5\% from SPF and the values
quoted above.

For comparison, in the semi-analytic model of Cole et al. (2000), only
5\% of the star formation at $z\sim0.7$ is contributed by the burst
mode (bursts were assumed to occur only in major mergers in this
model). In the updated model of Baugh et al. (2005), 16 \% of the star
formation at $z\sim0.7$ is from bursts \emph{of all types} (major +
minor). The new Baugh et al. model includes bursts in minor mergers,
but only when the gas fraction is above a critical value. Again, these
results only record the star formation during the relatively short
burst phase, and do not take into account the longer timescale of the
morphologically disturbed phase.

\section{Summary} 
 
We have investigated the contribution of galaxies of different 
morphological type to the UV luminosity density of the Universe at 
$z\sim 0.7$. The goal was to understand in which kind of galaxies the 
observed decline in the UV luminosity density occurred between this 
epoch and the present day. The experiment has shed light on the physical 
processes behind this drop in cosmic SFR. In particular, we have placed 
constraints on the fraction of the UV luminosity density arising from 
clearly interacting galaxies at $z\sim0.7$. 
 
Our analysis was based on a sample of 1483 galaxies in a thin redshift  
slice at $0.65<z<0.75$ in the COMBO-17/GEMS field encompassing the CDFS.  
We have used rest-frame luminosities at 280~nm and in the $V$-band from  
COMBO-17 together with three independent visual morphological  
classifications from rest-frame $V$-band images from the GOODS and GEMS  
HST/ACS surveys. We differentiated between four broad morphological types:  
spheroid-dominated (E--S0), disk-dominated (Sa--Sd), irregular (Irr) and  
peculiar or clearly interacting galaxies with morphological features  
indicative of ongoing major mergers (Pec-Int). The difference between Irr  
and Pec-Int is driven primarily by a desire to identify the cause of  
irregular morphological appearance; galaxies with morphologies consistent  
with the stochastic propagation of star formation were labeled irregular,  
while galaxies with features indicative of strong gravitational  
disturbance, such as multiple nuclei or tidal tails, were labeled Pec-Int. 
 
Comparing the morphological classifications between the shallower GEMS 
images and the deeper GOODS images, we have established that GEMS images 
do not allow for a complete identification of merging galaxies in the 
faintest magnitude bin we consider. This causes an artificially steep  
downturn in the luminosity function of mergers from the GEMS sample. 
Compensating for this shortcoming, the GEMS data offer the important  
advantage of a much larger galaxy sample with properties closer to the  
cosmic average, compared to GOODS. 
 
We analyzed UV-optical color-magnitude diagrams, UV luminosity functions  
and luminosity densities $j_{280}$ for all morphological types of galaxies.  
We found that although interacting and violently-merging galaxies dominate  
the rare group of particularly blue and very luminous objects, their  
rest-frame colors are similar to those of otherwise normal spirals.  
 
We found that seemingly normal disk galaxies (Sa--Sd) are the largest  
contributor to the UV luminosity density, emitting about half of the 
$j_{280}$ at $z\sim0.7$ and $M_{280}<-19$. The observed galaxy sample  
contributes 66\% of the total extrapolated $j_{280}$ from the entire  
galaxy population. After extrapolation, spiral galaxies emit $\sim$40\% 
of the total UV luminosity density. The second largest contribution  
originates from irregulars, which are predominantly faint and blue.  
Their steep luminosity function drives their contribution up to about  
40\% of the total $j_{280}$. Most of the remaining  
UV light comes from interacting systems and ongoing violent mergers,  
while the contribution from spheroids is negligible. The exact numbers 
for the merger contribution depend to some extent on the completeness of 
identification and the precise shape of their faint LF, but is likely to 
be around 20\%. We find that these results are consistent with predictions 
from a semianalytic galaxy formation model similar to the collisional 
starburst model of Somerville, Primack, \& Faber (2001). 
 
The results presented here demonstrate clearly that the drop in the UV 
luminosity density of the Universe since $z=1$ largely reflects decreasing  
SFRs in normal spirals. In parallel, high star formation  
rates in faint blue irregulars are migrating to systems of progressively  
lower mass and luminosity. Any evolution in the major merger rate plays  
only a small role in the decline of unobscured star formation. They could 
only be relevant for driving the decline of star formation since $z=1$  
if the drop in the cosmic SFR is substantially steeper than that inferred 
in the UV, and if the  
average extinction levels in mergers were far above the average  
of highly star-forming, but otherwise normal, spiral galaxies.   
First analyses of 24{\micron} data from Spitzer  
(Bell et al.\ 2005; Le Floc'h et al., in preparation) have started 
to address these issues, and support the conclusion of this  
paper that merging systems do not dominate the $z\la 1$ SFR.

\acknowledgements  
 
Support for the GEMS project was provided by NASA through grant number 
GO-9500 from the Space Telescope Science Institute, which is operated 
by the Association of Universities for Research in Astronomy, Inc. for 
NASA, under contract NAS5-26555. EFB and SFS ackowledge financial support 
provided through the European Community's Human Potential Program under 
contract HPRN-CT-2002-00316, SISCO (EFB) and HPRN-CT-2002-00305, Euro3D 
RTN (SFS). SJ acknowledges support from the National Aeronautics and 
Space Administration (NASA) under LTSA Grant NAG5-13063 issued through 
the Office of Space Science. DHM acknowledges support from the National 
Aeronautics and Space Administration (NASA) under LTSA Grant NAG5-13102 
issued through the Office of Space Science. CYP acknowledges support 
from NASA/JPL, under the Michelson Fellowship Program. JPL is managed 
for NASA by the California Institute of Technology. CH acknowledges 
support from the German-Israeli-Foundation, GIF. KJ was supported by the 
German DLR under project number 50~OR~0404. CW is supported by a PPARC 
Advanced Fellowship. We thank an anonymous referee for numerous helpful 
comments and Sadegh Khochfar for improving the manuscript.

\end{document}